\documentclass[aps,prd,twocolumn,
comment, showpacs,
superscriptaddress,groupedaddress,
amsmath,amssymb,floatfix,nofootinbib]
{revtex4}  
\usepackage{float}
\usepackage{graphicx}
\usepackage{bm}% bold math
\usepackage[usenames,dvipsnames]{color}
\usepackage{slashed}
\usepackage{hyperref}
\usepackage{slashed}

\begin{document}

%%%%
%    Greek Letters
%

\let\a=\alpha      \let\b=\beta       \let\c=\chi        \let\d=\delta
\let\e=\varepsilon \let\f=\varphi     \let\g=\gamma      \let\h=\eta
\let\k=\kappa      \let\l=\lambda     \let\m=\mu
\let\o=\omega      \let\r=\varrho     \let\s=\sigma
\let\t=\tau        \let\th=\vartheta  \let\y=\upsilon    \let\x=\xi
\let\z=\zeta       \let\io=\iota      \let\vp=\varpi     \let\ro=\rho
\let\ph=\phi       \let\ep=\epsilon   \let\te=\theta
\let\n=\nu
\let\D=\Delta   \let\F=\Phi    \let\G=\Gamma  \let\L=\Lambda
\let\O=\Omega   \let\P=\Pi     \let\Ps=\Psi   \let\Si=\Sigma
\let\Th=\Theta  \let\X=\Xi     \let\Y=\Upsilon
%
%%%

%%%
%    Calligraphic letters

\def\cA{{\cal A}}                \def\cB{{\cal B}}
\def\cC{{\cal C}}                \def\cD{{\cal D}}
\def\cE{{\cal E}}                \def\cF{{\cal F}}
\def\cG{{\cal G}}                \def\cH{{\cal H}}
\def\cI{{\cal I}}                \def\cJ{{\cal J}}
\def\cK{{\cal K}}                \def\cL{{\cal L}}
\def\cM{{\cal M}}                \def\cN{{\cal N}}
\def\cO{{\cal O}}                \def\cP{{\cal P}}
\def\cQ{{\cal Q}}                \def\cR{{\cal R}}
\def\cS{{\cal S}}                \def\cT{{\cal T}}
\def\cU{{\cal U}}                \def\cV{{\cal V}}
\def\cW{{\cal W}}                \def\cX{{\cal X}}
\def\cY{{\cal Y}}                \def\cZ{{\cal Z}}

%%%%

\def\be{\begin{equation}}
\def\ee{\end{equation}}
\def\bea{\begin{eqnarray}}
\def\eea{\end{eqnarray}}
\def\bm{\begin{matrix}}
\def\em{\end{matrix}}
\def\bpm{\begin{pmatrix}}
    \def\epm{\end{pmatrix}}

{\newcommand{\lsim}{\mbox{\raisebox{-.6ex}{~$\stackrel{<}{\sim}$~}}}
{\newcommand{\gsim}{\mbox{\raisebox{-.6ex}{~$\stackrel{>}{\sim}$~}}}
\def\mpl{M_{\rm {Pl}}}
\def\gev{{\rm \,Ge\kern-0.125em V}}
\def\tev{{\rm \,Te\kern-0.125em V}}
\def\mev{{\rm \,Me\kern-0.125em V}}
\def\ev{\,{\rm eV}}

\title{\boldmath   Explaining the CMS $eejj$ and $e \slashed {p}_T jj$ Excess and Leptogenesis in Superstring Inspired $E_6$ Models}
\author{Mansi Dhuria}
\email{mansi@prl.res.in}
\affiliation{Physical Research Laboratory, Navrangpura, Ahmedabad 380 009, India}
\author{Chandan Hati}
\email{chandan@prl.res.in} 
\affiliation{Physical Research Laboratory, Navrangpura, Ahmedabad 380 009, India}
\affiliation{Indian Institute of Technology Gandhinagar, Chandkheda, Ahmedabad 382 424, India.}
\author{Raghavan Rangarajan}
\email{raghavan@prl.res.in}
\affiliation{Physical Research Laboratory, Navrangpura, Ahmedabad 380 009, India}
\author{Utpal Sarkar}
\email{utpal@prl.res.in} 
\affiliation{Physical Research Laboratory, Navrangpura, Ahmedabad 380 009, India}

\begin{abstract}
	
	We show that superstring inspired $E_6$ models can explain both the recently detected excess $eejj$ and  $e \slashed p_T jj$ signals at CMS, and also allow for leptogenesis. Working in a R-parity conserving low energy supersymmetric effective model, we show that the excess CMS events can be produced via the decay of exotic sleptons in Alternative Left-Right Symmetric Model of $E_6$, which can also accommodate leptogenesis at a high scale. On the other hand, either the $eejj$ excess or the $e \slashed p_T jj$ excess can be produced via the decays of right handed gauge bosons, but some of these scenarios may not accommodate letptogenesis as there will be strong $B-L$ violation at low energy, which, along with the anomalous fast electroweak $B+L$ violation, will wash out all baryon asymmetry. Baryogenesis below the electroweak scale may then need to be implemented in these models.
	
\end{abstract}
%Particle-theory models (Early Universe), 98.80.Cq
%models beyond the standard models, 12.60.-i
%right-handed, 14.60.St
%Lepton number, 11.30.Fs
\maketitle

\section{Introduction}
Recently the CMS Collaboration at the LHC at CERN announced their results for the right-handed gauge boson $W_{R}$ search at a center of mass energy of $\sqrt{s}=8 \rm{TeV}$ and $19.7 \rm{fb}^{-1}$ of integrated luminosity. They have used the final state $eejj$ to probe $pp\rightarrow W_{R}\rightarrow eN_{R}\rightarrow eejj$, with the cuts $p_{T}> 60 \gev, \lvert \eta \rvert < 2.5$ ($p_{T} > 40 \gev, \lvert \eta \rvert < 2.5$) for leading (subleading) electron. The invariant mass $m_{eejj}$ is calculated for all events satisfying $m_{ee}>200 \gev$. In the bin $1.8 \tev< m_{eejj}<2.2 \tev$ roughly 14 events have been observed with 4 expected background events, amounting to a $2.8\sigma$ local excess, which, however, can not be explained by $W_R$ decay in Left-Right Symmetric Models (LRSM) with strict left-right symmetry (gauge couplings $g_L=g_R$) \cite{Khachatryan:2014dka}.  The CMS search for di-leptoquark production, at a center of mass energy of $\sqrt{s}=8 \rm{TeV}$ and $19.6 \rm{fb}^{-1}$ of integrated luminosity has reported a $2.4\sigma$ and a $2.6\sigma$ local excess in $eejj$ and $e\slashed{p}_{T}jj$ channels \footnote{The $e\slashed{p}_{T}jj$ channel is often referred to as $e\nu jj$ channel in the literature. Also note that the ``$ee$'' in $eejj$ refers to two first generation charged leptons, not necessarily of the same sign.} respectively, and has excluded the first generation scalar leptoquarks with masses less than 1005 (845) $\gev$ for $\beta =1(0.5)$, where $\beta$ is the branching fraction of a leptoquark to a charged lepton and a quark \cite{CMS:2014qpa}. In the $eejj$ channel for a $650 \gev$ leptoquark signal using the optimization cuts $S_{T} > 850\gev, m_{ee}>155\gev$ and $m^{\rm {min}}_{ej}>270\gev$ (where $S_{T}$ is the scalar  sum of the $p_{T}$ of two leptons and two jets), $36$ events have been observed compared with $20.49\pm 2.4\pm 2.45$(syst.) expected events from the Standard Model (SM) backgrounds implying a $2.4\sigma$ local excess. While in the $e\slashed{p}_{T}jj$ channel using the optimization cuts $S_{T} > 1040\gev, m_{\slashed{E}_{T}}>145\gev, m_{ej}>555\gev$ and $m_{T,e\slashed{p}_{T}}$ (where $S_{T}$ is now the scalar  sum of the missing energy $\slashed{E}_{T}$ and $p_{T}$ of the electron and two jets), $18$ events have been observed compared with $7.54\pm 1.20\pm 1.07$(syst.) expected background events amounting for a $2.6\sigma$ local excess.

 A few attempts have been made to explain the above CMS excesses in the context of different models. The excesses have been explained  in the context of $W_R$ decay by embedding the conventional LRSM $(g_L\ne g_R)$ in the $SO(10)$ gauge group in  Refs. \cite{Heikinheimo2014,Deppisch:2014qpa}. The $eejj$ excess has been discussed in the context of $W_R$ and $Z^\prime$ gauge boson production and decay in Ref. \cite{Saavedra2014}. The excesses have also been interpreted as due to pair production of vector-like leptons in Refs. \cite{Dobrescu2014}. In  Refs. \cite{Allanach-2014, Biswas-2014}, the excess of $eejj$ events has been shown to occur in $R$-parity violating  processes via the resonant production of a slepton. In Ref. \cite{strumia-2014}, a  different scenario is proposed by connecting leptoquarks  to dark matter which fits the data for the recent excess seen by CMS. The feasibility of probing lepton number violation through the production of same sign leptons pairs in a dilepton +2 jets channel was first explored in Ref. \cite{Keung:1983uu}.

The conventional LRSM (even embedded in higher gauge groups) are inconsistent with the canonical mechanism of leptogenesis in the  predicted range of the mass of $W_R$ $(\sim 2 \tev)$  at CMS \cite{Ma:1998sq}. In these models, leptogenesis can generate the lepton asymmetry in two possible ways: (i) decay of right-handed Majorana neutrinos which do not conserve lepton number \cite{Fukugita:1986hr} and (ii)  decay of very heavy Higgs triplet scalars with couplings that break lepton number \cite{Ma:1998dx}. Since the right-handed neutrinos interact with the $SU(2)_R$ gauge bosons, the $W_R$ interactions with the right-handed neutrino $N$ can wash out any existing primordial $B-L$ asymmetry, and hence also baryon and lepton asymmetry  in the presence of anomalous $(B+L)$ violating interactions before the electroweak phase transition. Successful primordial leptogenesis involving $N$ decay for $m_{W_R} > m_{N}$ then requires $m_{W_{R}}\gtrsim 2\times 10^{5} \gev (m_{N}/10^{2} \gev)^{3/4} $  if leptogenesis occurs at $T=m_{N}$ and $m_{W_{R}}\gtrsim 3\times 10^{6} \gev (m_{N}/10^{2} \gev)^{2/3}$ if it occurs at $T> m_{W_{R}}$. Note that the $m_{N} > m_{W_R}$ option is excluded in supersymmetric theories with gravitinos \cite{Ma:1998sq}. Thus, the observed $1.8 \tev < m_{W_{R}}<2.2 \tev $ bounds at CMS imply that the decay of right-handed neutrinos cannot generate the required amount of lepton and baryon asymmetry of the universe in the conventional LRSM. Such models would then require some other mechanism to generate the baryon asymmetry of the universe. With regards to the interpretation of the excess events as due to leptoquarks, it is difficult to accommodate any scenario that allows a light leptoquark in any simple extension of the standard model.

In this article, we attempt to find other simple extensions of the standard model which can explain the excess CMS events and simultaneously explain the baryon asymmetry of the universe via leptogenesis. To this end we explore whether models based on heterotic superstring theory have all the necessary ingredients embedded in their effective low-energy theories.  The heterotic superstring theory with $E_8 \times E'_8$ gauge group after compactification on a Calabi-Yau manifold leads to the breaking of $E_8
\rightarrow SU(3) \times E_6$  \cite{Candelasetal1985, greene}. The flux breaking of $E_6$ results in different effective subgroups of rank-5 and rank-6 at low-energy, some of which include new right-handed gauge bosons in their spectrum. In addition to this, these low-energy subgroups provide the existence of  new exotic (s)particles. We systematically study the possible decay modes of right-handed gauge bosons in the three effective low-energy subgroups of the superstring inspired $E_6$ model and point out that it is not possible to explain the excess of both $eejj$ and $e {\slashed p_T jj}$ events from the the right-handed gauge boson decay and accommodate leptogenesis simultaneously  in any of the effective low-energy subgroups of $E_6$.  We then propose a different scenario in which  both  the excess signals can be produced from the decay of an exotic slepton in two of the effective low-energy subgroups of the superstring inspired $E_6$ model. The added advantage of this scenario is that unlike R-parity violating slepton decay in Refs. \cite{Allanach-2014, Biswas-2014},  the production as well as decay of the exotic slepton in this scenario involves only R-parity conserving interactions. Interestingly,  one of the two effective low-energy subgroups (generally known as the Alternative Left-right Symmetric Model (ALRSM)) also explains high-scale leptogenesis. Therefore, we argue that the  ALRSM is the most suitable choice for  explaining both the excess of events at CMS and the generation of the baryon asymmetry via leptogenesis.

The outline of the article is as follows.  In Sec. {\bf II}, we discuss the phenomenology of the effective low-energy subgroups of the supersymmetric $E_6$ model and discuss the possibility of producing $eejj$ and $e{\slashed p_T jj}$ events from the decay of right-handed gauge bosons. In Sec. {\bf III}, we discuss the R-parity conserving processes for the production of an exotic slepton which subsequently decays to two jets and two electrons as well as two jets, an electron and missing energy. In Sec. {\bf IV}, we discuss
the impact of these models on leptogenesis. In Sec. {\bf V}, we conclude with our results.

\section{$E_6$-subgroups and decay channels of right-handed gauge bosons}
\label{sec_E6subgroups}

Within the context of heterotic superstring theory in ten
dimensions,  it was shown in Ref. \cite{GS1984} that there is
gauge and gravitational anomaly cancellation  if the underlying
gauge group is  $E_8 \times E_8'$ or $SO(32)$.  The $E_8 \times
E_8'$   leads to chiral fermions, whereas  $SO(32)$ does not lead
to the same.  Therefore, $E_8 \times E_8'$  is considered to be
more attractive from the phenomenological point of view.  By
integrating out  the massive modes, the low-energy limit of the
superstring theory (massless modes of the string) leads to
ten-dimensional supergravity coupled with an  $E_8 \times E_8'$
gauge sector. To make connection with the four-dimensional world,
the extra six dimensions must be compactified on a particular kind
of manifold. Though there exists several compactification
scenarios, the compactification on a Calabi-Yau manifold (with
SU(3) holonomy) \cite{greene} results in the breaking of $E_8
\rightarrow SU(3) \times E_6$  (with the $SU(3)$ gauge connection
becoming the spin connection on the compactified space) and also
produces ${\cal N}=1$ supersymmetry \cite{Candelasetal1985}.  The
remaining $E'_8$ couples to the usual matter representations of
the $E_6$ only by gravitational interactions and  provides the role
of the hidden sector needed to break supersymmetry.

Historically, the heterotic  superstring theory  made a lot of
progress because of its ability to produce ${\cal N}=1$
supersymmetry and its connection to the ${\cal N}=2$ superconformal field
theory allowing it to estimate the texture of the Yukawa couplings
in terms of intersection numbers in the compactified space.
Compactification in the Calabi-Yau space has another interesting
feature of flux breaking of the gauge groups at the grand
unification scale and also predicting three generations of
fermions. However, issues related to moduli stabilization as well as
the presence of exotic particles made it slightly less attractive.
On the flip side, the same exotic fermions appearing in
the $E_6$ spectrum provide a rich phenomenology and as we shall
demonstrate here, they may explain the LHC excess events. In this
section we discuss the phenomenology of the effective low-energy
subgroups of the $E_{6}$ group.  One of the maximal subgroups of
$E_{6}$ is given by $SU(3)_{C}\times SU(3)_{L} \times SU(3)_{R}$.
The fundamental $27$ representation of $E_{6}$ under this subgroup
decomposes as
\be{\label{4.0.0}} 27= (3, 3, 1)+(3^{\ast}, 1,
3^{\ast})+(1, 3^{\ast},3)
\ee
where $(u, d, h): (3, 3, 1)$, $(h^{c}, d^{c}, u^{c}): (3^{\ast},
1, 3^{\ast})$ and the leptons are assigned to $(1, 3^{\ast} ,3)$.
Here $h$ denotes the exotic $-\frac{1}{3}$ charge quark. Other
than $h$ and its charge conjugate, a right-handed neutrino $N^{c}$
and two lepton isodoublets $(\nu_{E}, E)$ and $(E^{c},N_{E}^{c})$
are among the new particles. Although these exotic particles have
not been observed so far, they promise rich phenomenology and
their detection may also become an indirect indication for the
superstring inspired models.

The particles of the first family are assigned as:
\be{\label{4.0}} \bpm u \\ d \\ h \epm + \bpm u^{c} & d^{c} &
h^{c}\epm +\bpm E^{c} & \nu & \nu_{E} \\ N^{c}_{E} & e & E \\
e^{c} & N^{c} & n \epm,
\ee
where $SU(3)_{L}$ operates vertically and $SU(3)_{(R)}$ operates
horizontally. When the $SU(3)_{(L,R)}$ further breaks to
$SU(2)_{(L,R)} \times U(1)_{(L,R)}$, there are three choices
corresponding to $T, U, V$ isospins of $SU(3)$, which corresponds
to the three different embedding of the residual $SU(2)$ on
$SU(3)$ and the different isospins $T, U, V$ would be the
generators of $SU(2)$. These three choices give three kinds of heavy $W$'s (compared to their left-handed counterparts) and the exotic fermions belong to the different $SU(2)$ representations.

\subsection{Case 1.}
We first consider the usual left-right symmetric extension of the
standard model and include the exotic particles. In that case, for
the standard model particles the sum of the generators $Y_L$ and
$Y_R$ can be identified with the generator $(B-L)$, where $B$ is
the baryon number and $L$ is the lepton number. We shall extend
this identification to the exotic particles as well, because that
will help us understand the $B$ and $L$ violation in this
scenario.

The right-handed up and down quarks, or their CP conjugate states
$(d^{c}, u^{c})_L$ belong to the $SU(2)_{R}$ doublet as in the LRSM.
The charge equation
 \bea Q&=&T_{3L}+\frac{1}{2} Y_{L}+T_{3R}+\frac{1}{2} Y_{R} \nonumber \\
&=&T_{3L} + T_{3R} +{(B-L) \over 2} 
\eea
holds for all the SM particles and we want the new fermions that
belong to the fundamental representation of $E_{6}$ to have
invariant Yukawa interactions with the SM particles. Thus this
relation may be extended as a definition to make all Yukawa and
gauge interactions conserve $B-L$. So under subgroup
$G=SU(3)_{c}\times SU(2)_{L}\times SU(2)_{R}\times U(1)_{B-L}$ the fields transform
as
 \bea {\label{4.1}}
(u, d)_{L} &:& (3, 2, 1, \frac{1}{6})\nonumber\\
(d^{c}, u^{c})_{L} &:& (\bar{3}, 1, 2, -\frac{1}{6})\nonumber\\
(\nu_{e}, e)_{L} &:& (1, 2, 1, -\frac{1}{2})\nonumber\\
(e^{c}, N^{c})_{L} &:& (1, 1, 2, \frac{1}{2})\nonumber\\
h_{L} &:& (3, 1, 1, -\frac{1}{3})\nonumber\\
h^{c}_{L} &:& (\bar{3}, 1, 1, \frac{1}{3})\nonumber\\
\bpm \nu_{E} & E^{c} \\ E & N^{c}_{E}\epm_{L} &:& (1, 2, 2, 0)\nonumber\\
n_{L} &:& (1, 1, 1, 0). \eea
 The presence of $SU(2)_{R}$ tells us that
the right-handed  charged currents must be incorporated in weak
decays. If the Dirac neutrino is formed combining $\nu_{e}$ and
$N^{c}$, then the mass of the $W_{R}^{\pm}$ is constrained from
polarized $\mu^{+}$ decay \cite{Bueno:2011fq}. Furthermore there
is a charged current mixing matrix for the known quarks in the
right-handed sector. Assuming this to be similar to Kobayashi-
Maskawa matrix one can constraint the $W_{R}^{\pm}$ mass from
the $K_{L}- K_{S}$ mass difference \cite{Beall:1981ze, Maiezza:2010ic,
Zhang:2007da}. In Ref. \cite{Senjanovic:2014pva} it was shown that the mixing matrix for the right-handed quark sector is calculable and that the difference between left and right mixing angles turns out to be very small. Also the kaon decay and neutron electric dipole moment can give further constraints on the $W_{R}$ mass \cite{Zhang:2007da, Ecker:1983dj}.

This case can produce the $eejj$ signal in the decays of $W_R$,
and can explain the observed events for $g_L \neq g_R$ \cite{Deppisch:2014qpa}. However,
this scenario is not very interesting for us as it cannot explain the
canonical mechanism of leptogensis as discussed in Sec. {\bf IV}.

\subsection{Case 2.}
Another choice for the $SU(2)_R$ doublet is $(h^{c}, u^{c})$  \cite{Ma:1986we} with
the charge equation $Q=T_{3L}+\frac{1}{2}
Y_{L}+T^{\prime}_{3R}+\frac{1}{2} Y^{\prime}_{R}$, where
\be {\label{4.2}} T^{\prime}_{3R}=\frac{1}{2} T_{3R}+\frac{3}{2}
Y_{R}, \ \ Y^{\prime}_{R}=\frac{1}{2} T_{3R}-\frac{1}{2} Y_{R},
\ee and we have $T^{\prime}_{3R}+Y^{\prime}_{R}=T_{3R}+Y_{R}$.
Thus it follows that for interactions involving only the standard
model particles and left-handed gauge bosons, one cannot distinguish between
the schemes of Case 1 and Case 2. In this scenario, the fields
transform under the subgroup $G=SU(3)_{c}\times SU(2)_{L}\times
SU(2)_{R^{\prime}}\times U(1)_{Y^{\prime}}$ as \bea {\label{4.3}}
(u, d)_{L} &:& (3, 2, 1, \frac{1}{6})\nonumber\\
(h^{c}, u^{c})_{L} &:& (\bar{3}, 1, 2, -\frac{1}{6})\nonumber\\
(\nu_{E}, E)_{L} &:& (1, 2, 1, -\frac{1}{2})\nonumber\\
(e^{c}, n)_{L} &:& (1, 1, 2, \frac{1}{2})\nonumber\\
h_{L} &:& (3, 1, 1, -\frac{1}{3})\nonumber\\
d^{c}_{L} &:& (\bar{3}, 1, 1, \frac{1}{3})\nonumber\\
\bpm \nu_{e} & E^{c} \\ e & N^{c}_{E}\epm_{L} &:& (1, 2, 2, 0)\nonumber\\
N^{c}_{L} &:& (1, 1, 1, 0),
\eea
 where $Y^{\prime}=Y_{L}+Y_{R}^{\prime}$. This model is often referred to as the Alternative Left Right
 Symmetric Model (ALRSM) in the literature. Note that, in this case $N^{c}$ has a
trivial transformation under G and thus can allow high-scale leptogenesis.
However, the assignment of quantum numbers for $N^{c}$ is not
unique and that can result in some consequences. 

With the above assignments, the superpotential governing interactions of Standard Model and exotic particles  is given as
 \begin{eqnarray}
 \label{eq:Wcase1}
 && W= \lambda_1\left( u u^{c} N^{c}_E - d u^{c} E^{c} - u h^{c} e + d h^{c} \nu_e \right)+ \nonumber\\
 &&  \lambda_2 \left( u d^{c} E - d d^{c} \nu_{E}\right)+\lambda_3 \left( h u^c e^c - h h^c n\right) + \nonumber\\
 && \lambda_4 h d^c N^{c}_L +\lambda_5 \left ( ee^c \nu_E + E E^c n - E e^{c} \nu_e- \nu_E N^{c}_E n\right) +\nonumber\\
 && \lambda_6 \left( \nu_e N^{c}_L N^{c}_E - e E^c N^{c}_L\right).
 \end{eqnarray}
 The superpotential given in Eq. (\ref{eq:Wcase1}) leads to the following assignments of $R$, $B$ and $L$ for the exotic fermions which also guarantees proton stability. For leptoquark $h$ we have $R=-1, B=\frac{1}{3}, L=1$; $\nu_{E}, E$ and $n$ carry $R=-1, B=L=0$. There are two possible assignments for $N^{c}$ determining whether a massive $\nu_{e}$ is possible or not. For the assignment $R=-1$ and $B=L=0$ for $N^{c}$ (which demands $\lambda_{4}=\lambda_{6}=0$ in Eq. (\ref{eq:Wcase1}) for a R-parity conserving scenario), one has an exactly massless $\nu_{e}$, but from the perspective of leptogenesis the more interesting choice is the case where $N^{c}$ is assigned $R=+1$, $B=0$, $L=-1$, so that it gives a tiny mass to $\nu_{e}$ via the seesaw mechanism. Thus we consider the latter scenario in the following discussions.

 In this case, the right-handed charged current couples $e$ to $n$, but with $n$ being presumably heavier ($m_{n} \gtrsim \cal O (\tev)$), there is no constraint on the mass of $W^{ \pm}_{R^{\prime}}$ from polarized $\mu^{+}$ decay in contrast to Case 1. Also since $W^{ \pm}_{R^{\prime}}$ does not couple to $d$ and $s$ quarks there is no constraint on the mass of  $W^{ \pm}_{R^{\prime}}$ from the $K_{L}- K_{S}$ mass difference either. Thus this model allows a much lighter $W_{R}^{\pm}$ than the Case 1. However this model can give rise to $D^{0}- \bar{D}^{0}$ mixing through the $W_{R^{\prime}}$ coupling of the $c$ and $u$ quarks to $h$ \cite{Ma:1987ji}. The relevant diagrams are shown in Fig. \ref{fig:wralrm2}.  This mixing can constrain the $SU(2)_{R^{\prime}}$ breaking scale in this model.
  % % % % % % % % % % % % % % % % % % % % % % % % % % % % % % % % % % % % % % % % % % % % % % % % % % % % % % % % % % % % % % % % % % % % % % % % % % % % % % % %
 \begin{figure}[h]
    \includegraphics[width=3.4in, height=1.8in]{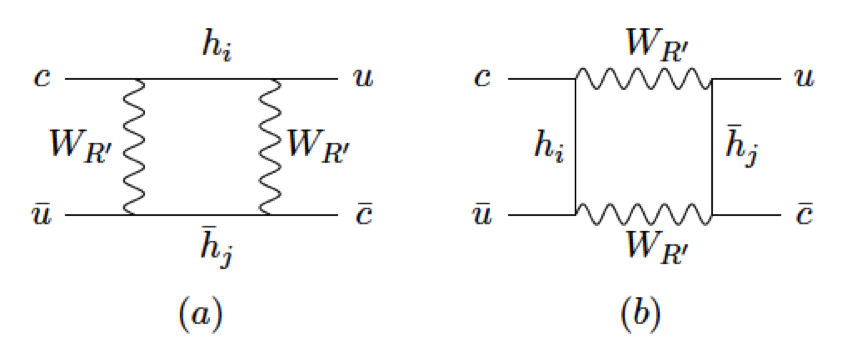}
    \caption{Box diagrams in the ALRSM  contributing to $D^{0}-{\bar D}^0$ mixing.}
    \label{fig:wralrm2}
 \end{figure}
 % % % % % % % % % % % % % % % % % % % % % % % % % % % % % % % % % % % % % % % % % % % % % % % % % % % % % % % % % % % % % % % % % % % % % % % % % % % % % % % %
It is interesting to note that in contrast to Case 1, where all the gauge bosons have $B=0$ and $L=0$, in this case $W^{ -}_{R^{\prime}}$ has leptonic charge $L=1$.  The coupling of the $W_{R^{\prime}}$ to the fermions is given by
\bea
\label{4.4}
&&{\hskip -0.3in}  \cL=\frac{1}{\sqrt{2}}  g_{R} W^{\mu}_{R^{\prime}} (\bar{h}^{c}\gamma_{\mu}u^{c}_{L} +\bar{E}^{c}\gamma_{\mu}\nu_{L}+\bar{e}^{c}\gamma_{\mu}n_{L}+\bar{N}^{c}_{E}\gamma_{\mu}e_{L})\nonumber\\
&&+ \rm{h.c.}
\eea
So  $W_{R^{\prime}}$ is coupled to the leptoquark $h^{c}_{L}$ and the $n$ field, compared to the coupling with the $d^{c}_{L}$ and $N^{c}$ in the conventional LRSM.

Let us discuss the possible production channels of $W_{R^{\prime}}$. The quantum numbers of $W_{R^{\prime}}$ imply that the production of $W_{R^{\prime}}$ from the usual $u\bar{d}$ scattering in hadronic colliders cannot take place. Furthermore since $2m_{W_{R^{\prime}}}> m_{Z^{\prime}}$, the pair production of $W_{R^{\prime}}$ via the decay of the heavy $Z^{\prime}$ is forbidden  \cite{Gunion:1987xi}. The process that can yield a large cross section for $W_{R^{\prime}}$ production is the associated production of $W_{R^{\prime}}$ and leptoquark via the process $g+u\rightarrow h+W_{R^{\prime}}^{+}$,
 which proceeds through the diagrams shown in Fig. \ref{fig:wralrm1}  \cite{Gunion:1987xi}.
\begin{figure}[h]
\includegraphics[width=3.4in, height=1.8in]{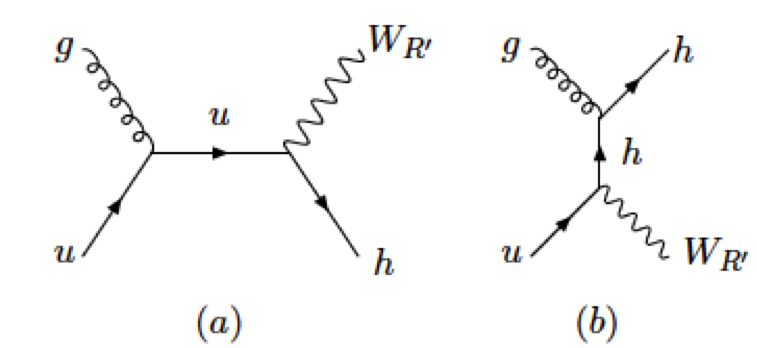}
    \caption{s- and t-channel Feynman diagrams for the process: $g+u \rightarrow  h+W_{R^{\prime}}$.}
    \label{fig:wralrm1}
\end{figure}
% % % % % % % % % % % % % % % % % % % % % % % % % % % % % % % % % % % % % % % % % % % % % % % % % % % % % % % % % % % % % % % % % % % % % % % % % % % % % % % %
The two-body decay modes of the $W_{R^{\prime}}$ can be obtained from Eq. (\ref{4.4}).  An inspection of all the further decays of the exotic particles coming from $W_{R^{\prime}}$ decay imply that the $W_{R^{\prime}}$ decay can not give rise to the  $ee+2j$ signal even in the presence of supersymmetry. However, there is a possibility to produce the $e\slashed p_T jj$ signal from the decay modes of $W_{R'}$ if `$n$' is considered as the Lightest Supersymmetric Particle (LSP). The relevant decay modes of $W_{R'}$ producing  $e\slashed p_T jj$ are given as: % %%%%%%%%%%%%%%%%%%%%%%%%%%%%%%%%%%%%%%%%%%%%%%%%%%%%%%%%%%%%%
%%%%%%%%%%%%%%%%%%%%%%%%%%%%%%%%%%%%%%%%%%%%%%%%%%%%%%%%%%%%%
  \bea {\label{4.6}}
 {\hskip -1.5in}& (i) &  W_{R^{\prime}}\rightarrow    {h}^{c}\bar {u^c} \rightarrow    {\tilde h}^{*} {\bar n} \bar {u^c} \rightarrow     \boxed{u^c e^c {\bar n} \bar {u^c}  }  \nonumber\\
&& {\hskip 0.75in}  \rightarrow     {\tilde e}^{*}  { \bar u}  \bar {u^c} \rightarrow    \boxed{{\bar e }{\tilde \gamma} { \bar u}  \bar {u^c}}\nonumber\\
&& {\hskip 0.75in}  \rightarrow      W_{R'} u^c \bar {u^c}  \rightarrow  \boxed{e^c {\bar n} u^c \bar {u^c}}  \nonumber\\
%\nonumber\\
%{\hskip -1.5in}& (ii) &  W_{R^{\prime}}\rightarrow    {E}^{c}{\bar \nu}_e \rightarrow     {u}^{c} \tilde{ e^c} u \rightarrow  \boxed{u^c \bar{e^c} {\tilde \gamma} u }\nonumber\\
%&& {\hskip 0.8in}  \rightarrow     { d}^{c} {\tilde \nu}_e u \rightarrow   \boxed{d^c {\tilde \gamma} \nu_e u}\nonumber\\
\nonumber\\
{\hskip -1.5in}& (iv) &  W_{R^{\prime}}\rightarrow    N^{c}_E {\bar e} \rightarrow   \bar {\nu_E} {\bar n} {\bar e} \rightarrow \boxed{d d^c {\bar n} {\bar e} }
 \eea
Thus, in this scenario, which has an attractive feature of allowing high-scale leptogenesis (as discussed in Sec. {\bf IV}),  a signal like two electrons and two jets can not correspond to the decay of $W_{R^{\prime}}$ whereas there are many channels which can produce a signal like  an electron, missing energy  and two jets via the decay of $W_{R^{\prime}}$ as given above. However, as we shall see in Sec. {\bf III},  both $eejj$ and $e\slashed p_Tjj$ signals can be analyzed in this case by considering $R$-parity conserving resonant production and decay of an exotic slepton.
%%%%%%%%%%%%%%%%%%%%%%%%%%%%%%%%%%%%%%%%%%%%%%%%%%%%%%%%%%%%%
% Presuming that mass of $n$ is smaller compared to $W_{R^{\prime}}, E$ and $N_{E}$ the kinematically dominant decay mode is expected to be $W^{-}_{R^{\prime}}\rightarrow e^{-}_{L}+n_{R}$.  Now if $n$ is the lightest SUSY particle (LSP) then the decay of $W_{R}$ gives $e\slashed{p}_{T}$, where $\slashed{p}_{T}$ denotes the missing ${p}_{T}$. If $n_{L}$ is not the LSP then it may decay into $\gamma \tilde{\gamma}$ to produce a signal $e \gamma \slashed{p}_{T}$. Even if $n_{R}$ is allowed to decay through an off-shell $W_{R^{\prime}}$: $n_{R} \rightarrow e^{+}+W^{-\ast}_{R^{\prime}}\rightarrow e^{+}+ u^{c}+h$, the final signal will depend on the decay modes of $h$. In a supersymmetric scenario the tree level decay modes of $h$ are
% % % % % % % % % % % % % % % % % % % % % % % % % % % % % % % % % % % % % % % % % % % % % % % % % % % % % % % % % % % % % % % % % % % % % % % % % % % % % % % %
  \subsection{Case 3.} A third way of choosing the $SU(2)_{R}$ doublet is $(h^{c}, d^{c})$ \cite{London:1986dk} and the charge equation is $$Q=T_{3L}+\frac{1}{2} Y_{L}+\frac{1}{2} Y_{N}\,,$$ where the $SU(2)$ corresponding to the mentioned doublet does not contribute to the electric charge equation and we will denote it as $SU(2)_{N}$. When this $SU(2)_{N}$ is broken, the gauge bosons $W^{\pm}_{N}$ and $Z_{N}$ acquires masses. Note that the $\pm$ in the superscript of $W_{N}$ refers to the $SU(2)_{N}$ charge. Under the subgroup $G=SU(3)_{c}\times SU(2)_{L}\times SU(2)_{N}\times U(1)_{Y}$ the fields transform as
%%%%%%%%%%%%%%%%%%%%%%%%%%%%%%%%%%%%%%%%%%%%%%%%%%%%%%%%%%%%%
\bea {\label{5.1}}
(u, d)_{L} &:& (3, 2, 1, \frac{1}{6})\nonumber\\
(h^{c}, d^{c})_{L} &:& (\bar{3}, 1, 2, \frac{1}{3})\nonumber\\
(E^{c}, N^{c}_{E})_{L} &:& (1, 2, 1, \frac{1}{2})\nonumber\\
(N^{c}, n)_{L} &:& (1, 1, 2, 0)\nonumber\\
h_{L} &:& (3, 1, 1, -\frac{1}{3})\nonumber\\
u^{c}_{L} &:& (\bar{3}, 1, 1, -\frac{2}{3})\nonumber\\
\bpm \nu_{e} & \nu_{E} \\ e & E\epm_{L} &:& (1, 2, 2, -\frac{1}{2})\nonumber\\
e^{c}_{L} &:& (1, 1, 1, 1).
\eea
%%%%%%%%%%%%%%%%%%%%%%%%%%%%%%%%%%%%%%%%%%%%%%%%%%%%%%%%%%%%%
 The superpotential governing interactions of SM and exotic particles  is given as:
 \begin{eqnarray}
 \label{eq:Wcase2}
 && {\hskip -0.2in} W= \lambda_1 \left( \nu_e N^{c}_L N^{c}_E + e E^c N^{c}_L +\nu_E N^{c}_E n + E E^c n \right) + \nonumber\\
 &&  {\hskip -0.2in} \lambda_2 \left( d^c N^{c}_L h+ h h^c n \right) +\lambda_3 u^c e^c h+ \lambda_4 \left( u u^c N^{c}_E + u^c d E^c\right)+ \nonumber\\
 &&  {\hskip -0.2in} \lambda_5 \left(\nu_e e^c E+ e e^c \nu_E \right) +\lambda_6 \left(u d^c E + d d^c \nu_E + u h^c e + d h^c \nu_e \right) \nonumber\\
  \end{eqnarray}
%%%%%%%%%%%%%%%%%%%%%%%%%%%%%%%%%%%%%%%%%%%%%%%%%%%%%%%%%%%%%
Note that in this case as well, the superpotential ensures that $h$ is a leptoquark ($B=\frac{1}{3}, L=1$) while $\nu_{E}, E$ and $n$ carry $B=L=0$ as in Case 2. $N^{c}$ has the assignment $B=0, L=-1$. $W_{N}$ has negative $R$-parity, nonzero leptonic charge $L=-1$ and zero baryonic charge.

 $W_{N}$ and $Z_{N}$ can induce $K^{0}- \bar{K}^{0}$ mixing. Consider a scenario where there is mixing between the six quarks (three generations) forming $SU(2)_{N}$ doublets
%%%%%%%%%%%%%%%%%%%%%%%%%%%%%%%%%%%%%%%%%%%%%%%%%%%%%%%%%%%%%
\be {\label{5.4}}
\bpm \bar{h_{1}} \\ \bar{d} \epm \ \ \bpm \bar{h_{2}} \\ \bar{s} \epm \\ \ \  \bpm \bar{h_{3}} \\ \bar{b} \epm
\ee
% % % % % % % % % % % % % % % % % % % % % % % % % % % % % % % % % % % % % % % % % % % % % % % % % % % % % % % % % % % % % % % % % % % % % % % % % % % % % % % % % % % % % % % % % % % % % %
\begin{figure}[h]
    \includegraphics[width=3.4in, height=1.8in]{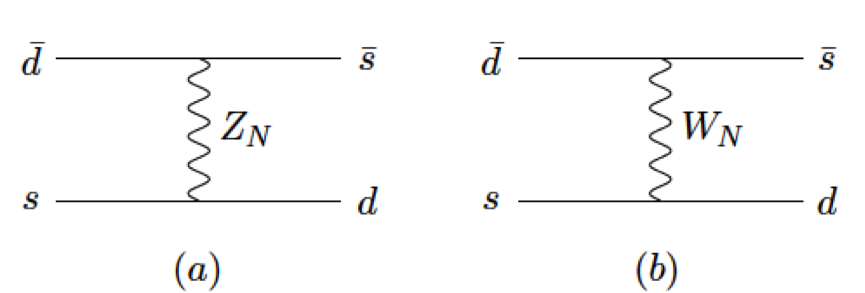}
    \caption{Tree level flavor changing neutral-current processes due to mixing of the six quarks, $d, s, b$ and
 exotic  quarks: $h_{i}$ ($i =1,2, 3$).}
    \label{fig:wn3}
\end{figure}
% % % % % % % % % % % % % % % % % % % % % % % % % % % % % % % % % % % % % % % % % % % % % % % % % % % % % % % % % % % % % % % % % % % % % % % % % % % % % % % % % % % % % % % % % % % % % %
\begin{figure}[h]
    \includegraphics[width=3.4in, height=1.8in]{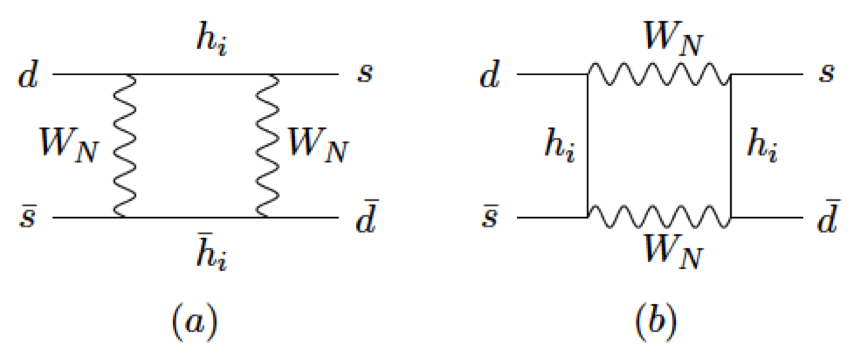}
    \caption{Box diagrams leading to $\bar{d}s-\bar{s}d$ mixing if only exotic $h_{i} (i=1,2,3)$ mix.}
    \label{fig:wn4}
\end{figure}
% % % % % % % % % % % % % % % % % % % % % % % % % % % % % % % % % % % % % % % % % % % % % % % % % % % % % % % % % % % % % % % % % % % % % % % % % % % % % % % %
Then the tree level Flavor Changing Neutral Current  (FCNC) processes as  shown in Fig. \ref{fig:wn3} will be present and one can get a bound for the $W_{N}$ from the $K_{L}-K_{S}$ mass difference \cite{London:1986dk}. Even if $\bar{d}$ and $\bar{s}$ do not mix with the exotic $\bar{h}_{i}$, there may still be a tree level contribution to the kaon mixing. If we assign opposite $T_{3N}$ to $\bar{d}_{L}$ and $\bar{s}_{L}$ and if they mix then the diagrams shown in Fig. \ref{fig:wn3} are still possible \cite{London:1986dk}. On the other hand if only the exotic $\bar{h}_{i}$ mix and the  $\bar{d}_{L}$ and $\bar{s}_{L}$ are assigned the same $T_{3N}$, then one gets the box diagrams shown in Fig. \ref{fig:wn4} \cite{London:1986dk}.  Similarly,  considering $SU(2)_{N}$ doublets in the leptonic sector,
%%%%%%%%%%%%%%%%%%%%%%%%%%%%%%%%%%%%%%%%%%%%%%
\be {\label{5.5}}
\bpm E \\ e \epm \ \  \bpm M \\ \mu \epm \ \ \bpm T \\ \tau \epm,
\ee
even in the absence of mixing between the ordinary and exotic fermions the process $\mu\rightarrow e \gamma$ can take place if the exotic fermions mix among themselves  \cite{London:1986dk} as shown in Fig. \ref{fig:wn5}.
% % % % % % % % % % % % % % % % % % % % % % % % % % % % % % % % % % % % % % % % % % % % % % % % % % % % % % % % % % % % % % % % % % % % % % % % % % % % % % % %
\begin{figure}[h]
    \includegraphics[width=3.1in, height=1.5in]{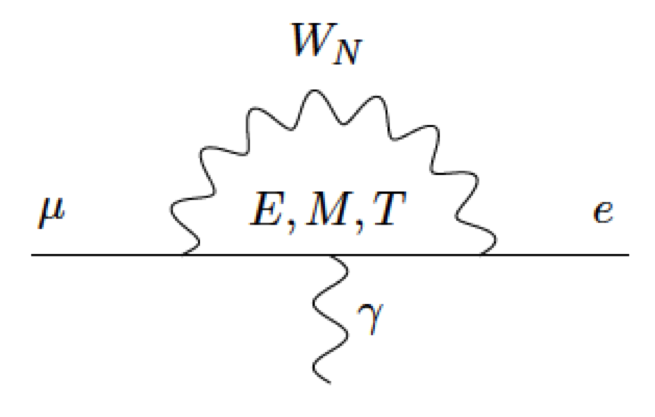}
    \caption{Loop diagrams involving exotic fermions and $W_{N}$ leading to $\mu \rightarrow e \gamma$.}
    \label{fig:wn5}
\end{figure}
The coupling of the $W_{N}$ to the fermions is given by
\be {\label{5.2}}
\cL=\frac{1}{\sqrt{2}}  g_{R} W^{\mu}_{N} (\bar{h}\gamma_{\mu}d_{R} +\bar{e}\gamma_{\mu} E_{L}+\bar{\nu}\gamma_{\mu}(\nu_{E})_{L}+\bar{N}^{c} \gamma_{\mu}n_{L})+ \rm{h.c.}
\ee
% % % % % % % % % % % % % % % % % % % % % % % % % % % % % % % % % % % % % % % % % % % % % % % % % % % % % % % % % % % % % % % % % % % % % % % % % % % % % % % %
\begin{figure}[h]
    \includegraphics[width=3in,height=1.5in]{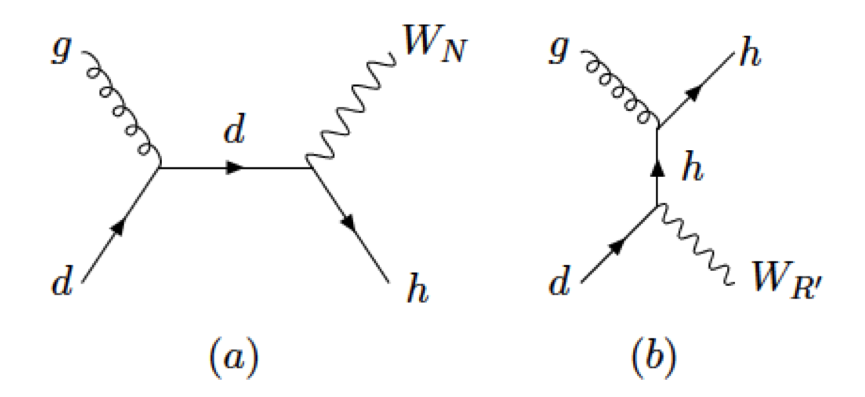}
    \caption{s- and t-channel Feynman diagrams for the process: $g+d \rightarrow  h+W_{N}$.}
    \label{fig:wn1}
\end{figure}
% % % % % % % % % % % % % % % % % % % % % % % % % % % % % % % % % % % % % % % % % % % % % % % % % % % % % % % % % % % % % % % % % % % % % % % % % % % % % % % %
On similar grounds, as in Case 2, the $W_{N}$ cannot be produced via the usual Drell-Yan mechanism or via the decay of the heavy $Z_{N}$. The process that can yield a large cross section for $W_{N}$ production \cite{Rizzo:1988bv} is $g+d\rightarrow h+W_{N}$ which consists of the diagrams shown in Fig. \ref{fig:wn1}. $W_{N}$ can also be produced in pairs via the process $e^{+}e^{-}\rightarrow W_{N}^{+} W_{N}^{-}$  \cite{Rizzo:1988bv}. The relevant diagrams are shown in Fig. \ref{fig:wn2}. As in the case of standard model $W^{\pm}$ pair production, this process is particularly sensitive to the gauge structure and cancellations between the contributing amplitudes. This provides a probe for non-Abelian $SU(2)_{N}$ gauge theory.
% % % % % % % % % % % % % % % % % % % % % % % % % % % % % % % % % % % % % % % % % % % % % % % % % % % % % % % % % % % % % % % % % % % % % % % % % % % % % % % %
\begin{figure}[h]
    \includegraphics[width=3in,height=1.6in]{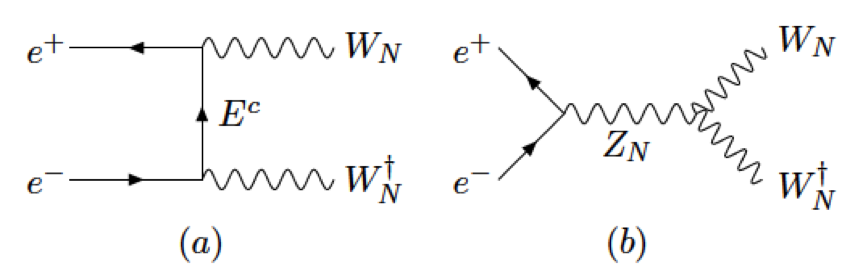}
    \caption{$s$- and $t$-channel Feynman diagrams for the process: $e^{+}e^{-} \rightarrow W_{N}^{+}W_{N}^{-}$.}
    \label{fig:wn2}
\end{figure}
% % % % % % % % % % % %
 Like in Case 2, an inspection of all the further decays of the exotic particles coming from the two-body decay modes of $W_{N}$ listed above tells us that a  $ee+2j$ signal cannot be obtained from the decay of $W_{N}$ even in the presence of supersymmetry.  However, there is a possibility to produce the $e\slashed p_T jj$ signal from the decay modes of $W_N$  if `$n$' is considered as Lightest Supersymmetric Particle (LSP). The relevant decay modes of $W_N$ producing  $e\slashed p_T jj$ are given as
   \bea {\label{4.6}}
 {\hskip -1.5in}& (i) &  W_{N}\rightarrow    {h}^{c}\bar {d^c} \rightarrow    {\tilde h}^{*} {\bar n} \bar {d^c} \rightarrow  \boxed{u^c e^c {\bar n} \bar {d^c}  }  \nonumber\\
&& {\hskip 0.7in}  \rightarrow     {\tilde e}^{*}  { \bar u}  \bar {d^c} \rightarrow      \boxed{{\bar e }{\tilde \gamma} { \bar u}  \bar {d^c}}\nonumber\\
\nonumber\\
{\hskip -1.5in}& (ii) &  W_N \rightarrow     e \bar{E_L} \rightarrow     e \tilde{E^c} { n}  \rightarrow \boxed{e \bar{u^c}  \bar{d} n}    \eea
%%%%%%%%%%%%%%%%%%%%%%%%%%%%%%%%%
Thus, similar to the Case 2,  a signal like two electrons and two jets can not correspond to the decay of $W_{N}$  whereas there are some channels which can produce a signal like  an electron, missing energy  and two jets via the decay of $W_{N}$ as given above. However, as shown in Sec. {\bf III}, both $eejj$ and $e\slashed p_Tjj$ signals can be interpreted in this case also by considering $R$-parity conserving resonant production and decay of an exotic slepton.
% % % % % % % % % % % % % % % % % % % % % % % % % % % % % % % % % % % % % % % % % % % % % % % % % % % % % % % % % % % % % % % % % % % % % % % % % % % % % % % %
\section{Exotic sparticle(s) production leading to an $eejj(e\slashed p_Tjj)$ signal}
In this Section, we show that two of the effective low-energy subgroups (discussed as `Case 2' and `Case 3' in Sec. {\bf II}) of the $E_6$ group can produce both $eejj$ and $e\slashed p_T jj$ signals from the decay of scalar superpartner(s) of the exotic particle(s). Both events can be produced naturally in the above schemes by considering (i) resonant production of the exotic slepton $\tilde E$ (ii) pair production of scalar leptoquarks ${\tilde h}$. As discussed in Sec. {\bf I}, CMS has already performed a search for the excess of both $eejj$ and $e\slashed p_Tjj$ events in the final states via pair production of leptoquarks. Interestingly, as compared to the resonant production of sleptons as discussed in Ref. \cite{Biswas-2014},  the exotic slepton ${\tilde E}$ can be resonantly produced in $pp$ collisions without violating $R$-parity. The exotic slepton then subsequently decays to a charged lepton and neutrino, followed by  R-parity conserving
interactions of the neutrino producing an excess of events in both  $eejj$ and $e\slashed p_Tjj$ channels. The R-parity conserving processes leading to both $eejj$ and $e\slashed p_T jj$ signals are given in Fig. \ref{fig:eejj2}(a) and the one giving only $eejj$ signal is given in Fig. \ref{fig:eejj2}(b).
 The cross section of the $eejj$ process as given in Fig. \ref{fig:eejj2}(a) and Fig. \ref{fig:eejj2}(b) can be expressed as:
 \be
 \sigma\left(pp \rightarrow eejj\right)= \sigma (pp \rightarrow {\tilde E_L}) \times BR(\tilde E_L \rightarrow eejj)
 \ee
  whereas the cross section of the $e\slashed p_Tjj$ process as given in Fig. \ref{fig:eejj2}(b) can be expressed as
  \bea
 \sigma\left(pp \rightarrow  e\slashed p_Tjj\right)= \sigma (pp \rightarrow {\tilde E_L}) \times BR(\tilde E_L \rightarrow e \slashed p_T jj).\nonumber\\
 \eea
 \begin{figure}
 \includegraphics[height=4cm,width=9cm]{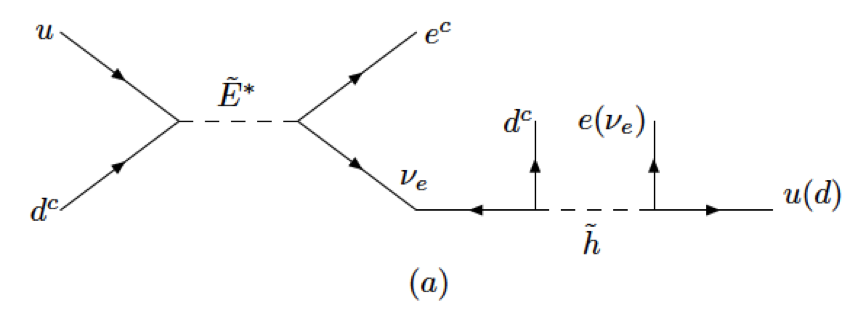}
\includegraphics[height=4cm,width=9cm]{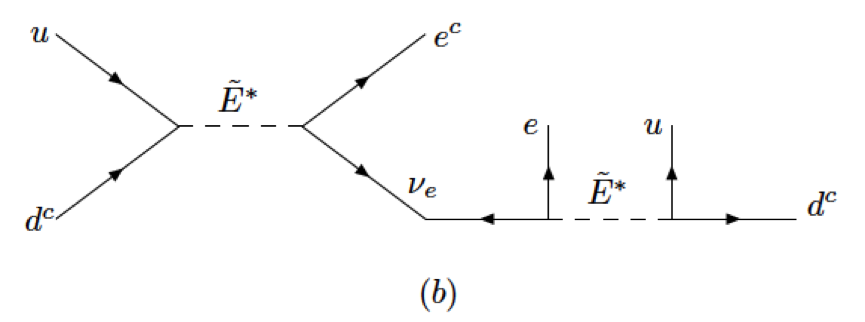}
\caption{R-parity conserving Feynman diagrams for a  single exotic particle ${\tilde E}$ production leading to both $eejj$ and $e {\slashed p_T}jj$ signal in Fig. 8(a) and $eejj$ signal in Fig. 8(b).}
 \label{fig:eejj2}
 \end{figure}
In Case 2, the resonant production of the slepton as well as decay modes of the same are  given by the following terms in the superpotential
 \be
 W_2=- \lambda_1\left(u h^{c} e - d h^{c} \nu_e \right)+    \lambda_2  u d^{c} E  -\lambda_5  E e^{c} \nu_e,
 \ee
while in Case 3, the relevant interaction terms are given by
 \be
 W_3 =  \lambda'_5 \left(\nu_e e^c E+ e e^c \nu_E \right) +\lambda'_6 \left(u d^c E + u h^c e + d h^c \nu_e \right).\nonumber\\
  \ee
   The  parton cross section of a single slepton production in Case 2 is given by \cite{Allanach-2009}
 \be
  {\hat \sigma}=\frac{\pi}{ 12 \hat s} \left| \lambda_{2}\right|^2 \delta (1-\frac{ m^{2}_{\tilde E}}{\hat s})
 \ee
 where ${\hat s}$ is the partonic centre of mass energy, and $m_{\tilde E}$ is the mass of the resonant slepton. Including effects from parton distribution functions,  the total cross section to a good approximation is given by \cite{Allanach-2009}
 \be
  \sigma\left(pp \rightarrow eejj\right) \propto \frac{\left | \lambda_{2}\right|^2}{m^{3}_{\tilde E}}  \times \beta_1  \ee
  and
  \be
  \sigma\left(pp \rightarrow e {\slashed p_T}j j\right) \propto \frac{\left | \lambda_{2}\right|^2}{m^{3}_{\tilde E}}  \times  \beta_2 
   \ee
where $\beta_1$ is the branching fraction for the decay of the exotic slepton to $eejj$ and $\beta_2$ is the branching fraction to $e\slashed p_T jj$. Similarly, in Case 3, the cross sections in the $eejj$ and $e\slashed p_T jj$ channels depend on $\frac{\left | \lambda_{6}\right|^2}{m^{3}_{\tilde E}}  \times \beta_1$ and  $\frac{\left | \lambda_{6}\right|^2}{m^{3}_{\tilde E}}  \times \beta_2$ respectively. By choosing $\beta_{1,2}$ as well as couplings $\lambda_{2} (\lambda'_{6})$ as free parameters in Case 2 (Case 3),  the cross section can be calculated as a function of the exotic slepton mass. Stringent bounds can also be obtained on the value of the mass of the exotic slepton by comparing the theoretically calculated cross section with the data collected by CMS at a centre of mass energy $\sqrt{s}=8$ TeV. Thus, we propose that the alternative schemes of $E_6$  might explain the excess   $eejj$ and $e {\slashed p_T}jj$ signals at CMS naturally via resonant  exotic slepton decay.
%%%%%%%%%%%%%%%%%%%%%%%%%%%%%%%%%%%%%%%%%%%%%%%%%%%%%%%%%%%%%%%%%%%%%%%%%%%%%%%%%%%%%%%%%%%%%%%%%%%%%%%%%%%%%%%%%%%%%%%%%
 \section{ Leptogenesis in supersymmetric low energy $E_6$-subgroups}
 	In the conventional LRSM scenario for  successful high-scale leptogenesis, constraints on the $W_{R}$ mass mentioned in the introduction follow from the out-of-equilibrium condition of the scattering processes involving the $SU(2)_{R}$ gauge interactions \cite{Ma:1998sq}. In the case $M_{N}>M_{W_{R}}$ the condition that the process
 	\be
 	e_{R}^{-}+W_{R}^{+}\rightarrow N_{R} \rightarrow e_{R}^{+}+W_{R}^{-}
 	\ee
 	goes out of equilibrium gives 
 	\be
 	M_{N}\gsim 10^{16} \gev
 	\ee
 	with $m_{W_{R}}/m_{N}\gtrsim 0.1$. For the case $M_{W_{R}}>M_{N} $ leptogenesis can occur either at $T\simeq M_{N}$ or at $T>M_{W_{R}}$ below the $B-L$ breaking scale. For $T\simeq M_{N}$, the out-of-equilibrium condition of the scattering processes which maintain the equilibrium number density for $N_R$ leads to
 	\be
 	M_{W_{R}}\gtrsim 2\times 10^{5} \gev (M_{N}/10^{2} \gev)^{3/4}.
 	\ee
 	For leptogenesis at $T>M_{W_{R}}$ the condition that the scattering process
 	\be
 	W_{R}^{\pm}+W_{R}^{\pm}\rightarrow e_{R}^{\pm} + e_{R}^{\pm}
 	\ee
 	through $N_{R}$ exchange goes out of equilibrium gives
 	\be
 	M_{W_{R}}\gtrsim 3\times 10^{6} \gev (M_{N}/10^{2} \gev)^{2/3}.
 	\ee
 	Consequently, observing a $W_{R}$ signal in the range $1.8 \tev < M_{W_{R}}<2.2 \tev $ implies that it is not possible to generate the required baryon asymmetry of the universe from the high-scale leptogenesis for a hierarchical neutrino mass spectrum ($M_{N_{3R}}\gg M_{N_{2R}}\gg M_{N_{1R}}=m_{N}$) in the usual LRSM scenario (even if it is embedded in a higher gauge group). 
 
The $E_6$ group allows the possibility of explaining leptogenesis in two of its effective low-energy subgroups  \cite{ma_2000}. One of them is $G_{1}=SU(3)_{C}\times SU(2)_{L}\times U(1)_{Y}\times U(1)_{N}$  and the other is $G_{2}=SU(3)_{C}\times SU(2)_{L}\times SU(2)_{R^{\prime}}\times U(1)_{Y^{\prime}}$, which is the Case 2 of Sec. {\bf II}. With the assignment given in Eqs. (\ref{4.1}), (\ref{4.3}) and (\ref{5.1}), amongst the five neutral fermions, only $\nu_{e}$ and $N^{c}$ carry nonzero $B-L$ in all subgroups. Therefore, leptogenesis can be addressed via the decay of the Majorana neutrino $N^c$ in all three cases.  Now to generate the $B-L$ asymmetry from the heavy neutrinos, one needs to satisfy the conditions: (i) violation of $B-L$ from Majorana mass of $N$ (ii) CP violation from complex couplings and (iii) the out-of-equilibrium condition of the decay rate of the physical heavy Majorana neutrino $N$ given by
\be{\label{4.1.1}}
\Gamma_{N}< H(T=m_{N})=\sqrt{\frac{4\pi^{3}g_{\ast}}{45}} \frac{T^{2}}{M_{Pl}},
\ee
where $\Gamma_{N}$ is the  decay width, $H(T)$ is the Hubble expansion rate, $g_{\ast}$ is the number of relativistic degrees of freedom at temperature $T$ and $M_{Pl}$ is the Planck mass. This translates into the condition that the mass of $N$ must be many orders of magnitude greater than the $\tev$ scale and consequently $N^{c}$ cannot transform nontrivially under the low-energy subgroup G. From the assignments of Eq. (\ref{4.1}) and Eq. (\ref{5.1}), it follows that $N^{c}$ transforms as a doublet under both $SU(2)_{R}$ and $SU(2)_{N}$. This implies that if $SU(2)_{R}$ gets broken at the $\tev$ scale, a successful leptogenesis scenario can not be obtained in Case 1 (similar to the conventional left-right model) and Case 3. In Case 2, since $N^c$ transforms trivially under the low-energy subgroup G,  the out-of-equilibrium decay of heavy neutrinos can give rise to high-scale leptogenesis \footnote{Note that  Ref. \cite{Hambye:2005tk} considers a scenario in which the lepton asymmetry is generated via Higgs triplet decays while the wash out processes involving gauge interactions are in effect. In this scheme, the leptogenesis can work out in the strong wash out regime. However in our case where lepton asymmetry gets generated at a high scale, the wash out processes involving gauge interactions (effective at a lower energy scale) must go out of equilibrium so that the lepton asymmetry does not get wiped out.}. 
In this case, the Majorana neutrino $N^c_{k}$ decays to $B-L=-1$ final states $\nu_{e_{i}} {\tilde N}^{c}_{E_{j}}, {\tilde \nu}_{e_{i}} N^{c}_{E_{j}}, e_{i}{\tilde E}^{c}_{j}, {\tilde e}_{i}, E^{c}_{j}$ and $d_{i} {\tilde h}_{j}, {\tilde d^{c}}_{i} {\tilde h}_{j}$ and to their conjugate states, via the interaction terms $\lambda_{4}$ and $\lambda_{6}$ in Eq. (\ref{eq:Wcase1}). One-loop diagrams, such as the two shown in Fig. \ref{fig:lg} for a given final state, can interfere with the tree level $N_{k}$ decays to provide the required CP violation for particular values of couplings $\lambda^{ijk}_{4}$ and $\lambda^{ijk}_{6}$ .
\begin{figure}[h]
    \includegraphics[width=3.4in,height=1.2in]{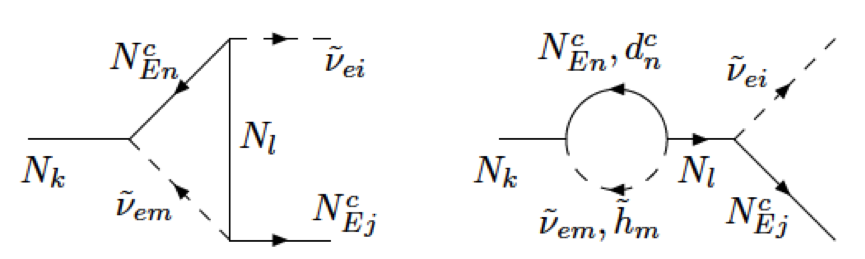}
    \caption{Loop diagrams for $N_k$ decay.}
    \label{fig:lg}
\end{figure}
An order of magnitude estimate of the upper bound on the couplings $\lambda^{ijk}_{4}$ and $\lambda^{ijk}_{6}$ can be obtained from the out-of-equilibrium condition given by Eq. (\ref{4.1.1}). Considering the total decay width of $N_{k}$ given by
\be{\label{4.1.2}}
\Gamma_{N_{k}}=\frac{1}{4\pi} \sum_{i,j}  (|\lambda^{ijk}_{4}|^{2} + 2|\lambda^{ijk}_{6}|^{2})m_{N_{k}}
\ee
and taking $g_{\ast}\sim 100$ at $T\sim m_{N_{k}}$, the condition given by Eq. (\ref{4.1.1}) gives 
\be{\label{4.1.3}}
 \sum_{i,j}  (|\lambda^{ijk}_{4}|^{2} + 2|\lambda^{ijk}_{6}|^{2})\lsim 2\times 10^{-17} \gev^{-1} m_{N_{k}}.
\ee
So for $m_{N_{k}}\sim 10^{15} \gev$, $\lambda^{ijk}_{4}, \lambda^{ijk}_{6}\lsim 10^{-1}$. In fact $\lambda^{ijk}_{4,6}\sim 10^{-3}$ can give the observed baryon-to-entropy ratio $n_{B}/s\sim 10^{-10}$ for maximal CP violation \cite{ma_2000}. Therefore, Case 2 (or ALRSM) has the attractive feature that it can explain both the excess $eejj$ and $e\slashed p_T jj$ signals, as shown in Sec. {\bf III}, and also high-scale leptogenesis.

\section{Conclusions}
We have studied the effective low-energy theories of the superstring inspired $E_6$ group that
is broken to its maximal subgroup by flux breaking at a very high scale. Our aim was to look
for extensions of the standard model that can explain the excess $eejj$  and  $e \slashed p_T jj$ 
events that have been observed by CMS at the LHC at the center of mass
energy $\sqrt{s} = 8$ TeV, and simultaneously explain
the baryon asymmetry of the universe via leptogenesis.

We consider generating the excess signals via both right-handed gauge boson decay and exotic slepton  decay. The decay of the right-handed gauge boson is able to produce $eejj$ events in one of the effective low-energy subgroups given by $G=SU(3)_c \times SU(2)_L \times SU(2)_R \times U(1)_{B-L}$ (Case 1 of Sec. {\bf \ref{sec_E6subgroups}}).  The  right-handed gauge boson decay in 
the other two effective low-energy subgroups of $E_6$: $G=SU(3)_c \times SU(2)_L \times SU(2)_{R'} \times U(1)_{Y'}$ (Case 2) and  $G=SU(3)_c \times SU(2)_L \times SU(2)_{N} \times U(1)_{Y}$ (Case 3), can produce the $e {\slashed p_T} jj$ signal if the exotic particle `$n$' is considered to be the LSP. However, none of these subgroups is  able to produce both excess 
signals simultaneously from the decay of right-handed gauge bosons. 
On the other hand, both signals can be produced simultaneously in the latter two effective low-energy subgroups of $E_6$ 
via the $R$-parity conserving resonant production of an exotic slepton, followed by its decay via $R$-parity conserving interactions. 
Now, since the effective low-energy subgroups of $E_6$ in Case 1
allow breaking of $U(1)_{B-L}$ at a scale lower than the $SU(2)_R$ breaking scale, it  is not consistent with leptogenesis at a high scale. The other two subgroups (Case 2 and Case 3) allow breaking of $SU(2)_{R}$ at a low scale which is independent of the $B-L$ breaking scale. However, since in Case 3, the right-handed neutrino transforms nontrivially under the low energy group, it can not give rise to high scale leptogenesis. In Case 2, since the the right-handed neutrino transforms trivially under the low energy group, it can allow for leptogenesis at a high scale. Therefore, we conclude that $G=SU(3)_c \times SU(2)_L \times SU(2)_{R'} \times U(1)_{Y'}$ subgroup, also referred to as the Alternative Left-Right Symmetric Model, can explain both the excess $eejj$ and $e {\slashed p_T} jj$ signals and also satisfy the constraints for successful leptogenesis.

%Though the CMS data shows  small deviation from the Standard Model, there is still some regime in the parameter space to explain the excess with the models  which include supersymmetry.
If future runs at LHC confirm the excess signals discussed above, it would be interesting to rely on the low-energy spectrum of $E_6$ group because it provides an approach to explain the excess of events while satisfying the constraints of successful leptogenesis. However, it would be challenging to  dynamically obtain the relevant Yukawa couplings in the context of the low-energy limit of $E_8 \times E'_8$ superstring theory for the theoretical cross section to match the data observed by CMS.
% at a centre of mass energy $\sqrt{s}=8$ TeV.

 % % % % % % % % % % % % % % % % % % % % % % % % % % % % % % % % % % % % % % % % % % % % % % % % % % % % % % % % % % % % % % % % % % % % % % % % % % % % % % % % % % % % % % % % % % % % % % % % % % % % % % % % % % % % % % % % % % % % % % % % % % % % % % % % % % % % % % % % % % % % % % % % % % % % % % % % % % % % % % % % % % % % % % % % % % % % % % % % % % % % % % % % % % % % % % % % % % % % % % % % % % % % % % % % % % % % % % % % % % % % %
 
\end{document}